\documentclass[aps,prb,preprint,showpacs]{revtex4}
\usepackage{bm}
\usepackage{graphicx}
\begin{document}
\title{Resonance-like electrical control of electron spin
for microwave measurement \protect\footnote{The following article has been submitted to 
Applied Physics Letters. After it is published, it will be found at 
http://apl.aip.org} }
\author{B.~A.~Glavin}
\affiliation{V.E.~Lashkar'ov Institute of Semiconductor Physics,
Pr.~Nauki 41, Kiev 03028, Ukraine}
\author{K.~W.~Kim}
\affiliation{Department of Electrical and Computer Engineering,
North Carolina State University, Raleigh, NC 27695-7911, USA}
\begin{abstract}
We demonstrate that the spin-polarized electron current can
interact with a microwave electric field in a resonant manner. The
spin-orbit interaction gives rise to an effective magnetic field
proportional to the electric current. In the presence of both dc
and ac electric field components, electron spin resonance occurs
if the ac frequency matches with the spin precession frequency
that is controlled by the dc field. In a device consisting of
two spin-polarized contacts connected by a two-dimensional
channel, this mechanism allows electrically tuned detection of the
ac signal frequency and amplitude. For GaAs, such detection is
effective in the frequency domain around tens of gigahertz.
\end{abstract}


\maketitle

During the last decade, there has been considerable interest in
spin-based applications for data storage and processing. One of
the widely discussed designs is the so-called spin field-effect
transistor.\cite{dasdatta} In this device, the source and drain
contacts are spin-polarized (which can be achieved, for example,
in a diluted magnetic semiconductor in the ferromagnetic phase).
Spin-polarized electrons injected from the source contact are
transferred to the drain through a quantum well (QW) channel. In
the course of this transfer, the electron spin precesses due to
the spin splitting of QW energy band spectrum. When the spin
splitting is mainly due to the Bychkov-Rashba
mechanism,\cite{byra} the magnitude of splitting and,
consequently, the frequency of spin precession can be modified by
the transverse electric field controlled by the gate voltage.
Therefore, both the electron spin polarization and current at the
drain are controlled by the gate voltage.

In this letter, we propose a different, gate-less scheme of
nonequilibrium spin control in a similar device, whose schematics
is shown in the insert of Fig.~1. The proposed method relies on
the interaction of spin-polarized electrons with a microwave
electric field. Although this interaction looks similar to the
electric-dipole resonance,\cite{rashe} the main difference is that
the effective splitting in our case is induced by the electric
current rather than the magnetic field and, therefore, allows
purely electrical tuning. We derive criteria for effective spin
control and analyze them for various semiconductor materials. Our
results suggest that the proposed device structure can be used to
determine the frequency, amplitude, and polarization of microwave
radiation with potential advantages over the existing
approaches.\cite{muwave}

Let us start with the consideration of spin dynamics in a QW in
the presence of electric field  ${\bm E}$. The electron
Hamiltonian is
\begin{equation}
\label{eq:1} H = \frac{\hbar^2 k^2}{2 m} + \frac{\hbar}{2} {\bm
\sigma\bm \Omega} + {\hat V} ~,
\end{equation}
where $k$ is the electron momentum and $m$ is the electron
effective mass.  The second term describes the spin splitting of
the electron spectrum in a $k$-dependent effective magnetic field
and the third term corresponds to the electron interaction with
the electric field. Two contributions to $\bm \Omega$ proportional
to the electron wavevector can be important. The bulk-induced
contribution  ${\bm \Omega}_b$  is due to the absence of the
inversion center in the crystal constituting the QW, while the
structure-induced contribution ${\bm \Omega}_s$  arises from the
asymmetry of the QW structure.\cite{byra} For the particular case
of a cubic QW grown along the (001) direction, we have ${\bm
\Omega}_b = \alpha (k_x; -k_y; 0)$ and ${\bm \Omega}_s = \beta
(k_y; -k_x; 0)$.  Here, $\alpha$ and $\beta$ are the material and
structure-dependent parameters. Since $\bm \Omega$ is an odd
function of the electron wavevector, its average value is zero if
there is no current in the system. For a finite current, however,
it becomes non-zero and the electrons experience the influence of
{\em average} effective magnetic field.\cite{previous} Here, we
make use of the fact that if both dc and ac components of the
electric current are present, they can give rise to electron spin
resonance even in the absence of ac and dc magnetic fields.

It is convenient to treat this resonance phenomenon by using a
hydrodynamic spin transport model.  If the electron density is
spatially uniform, the evolution of the average electron spin $\bm
S$ can be described by the following equation:\cite{wig}
\begin{equation}
\label{eq:2}
\frac{\partial {\bm S}}{\partial t} =
\bar{\bm \omega}\times {\bm S} - v_x \frac{\partial {\bm S}}{\partial x}
+D \frac{\partial^2{\bm S}}{\partial x^2} -{\hat \gamma} {\bm S}.
\end{equation}
Here, we assume that the electric current is parallel to the $x$
axis; $v_x$ is the $x$-component of the electron drift velocity,
$D$ is the diffusion coefficient, $\bar{\bm \omega}$ is the {\em
average} effective magnetic field in units of frequency, and $\hat
\gamma$ is the relaxation matrix. Equation~(\ref{eq:2}) is
applicable if all the characteristic times related to spin
evolution are much longer than the electron orbital relaxation
times or, in other words, when the spin-orbit (SO) interaction is
weak. To be specific,  we consider in the following the case when
the bulk-induced contribution is the most important [i.e.,
$\bar{\bm \omega} = \alpha (m v_x/\hbar ; -m v_y/\hbar; 0)$] and
spin relaxation is due to the D'yakonov-Perel' mechanism.\cite{DP}
These conditions correspond, in particular, to GaAs/AlGaAs QWs. In
this case, the nonzero components of the relaxation matrix are
$\gamma_{xx}=\gamma_{yy}=\gamma$, $\gamma_{zz} = 2 \gamma$,
$\gamma \approx (\Omega ({\bar k}))^2 \tau$, where $\bar k$ is the
characteristic electron wavevector and $\tau$ is the electron
relaxation time.\cite{DP}

In the following, we neglect spin diffusion (the justification
will be discussed explicitly below) and assume that the dc (ac)
field is parallel to the $x$ ($y$) axis. Under these conditions,
$\bar{\bm \omega} = {\bm \omega_0} + {\bm \omega_1} \cos \omega
t$, where ${\bm \omega}_0 = \alpha (m v_0/\hbar ; 0; 0)$, ${\bm
\omega}_1 = \alpha (0; -m v_1 /\hbar; 0)$, $v_0$ and $v_1 \cos
\omega t$ are the dc and ac drift velocities, respectively, and
$\omega$ is the frequency of the ac signal. For convenience, we
define $ \alpha m v_0/\hbar $ and $ \alpha m v_1/\hbar $ as $
\omega_0 $ and $ \omega_1$, respectively. The growth direction is
chosen to be the $z$ [(001)] axis.

The solution of Eq.~(\ref{eq:2}) with $D=0$ can be sought in the
form ${\bm S} (x,t) = {\bm S}(x-v_0 t,t)$, which transforms it
into the ordinary differential equations similar to the Bloch
equations for spin resonance.\cite{abragam} If the initial spin
state near the source ($x=0$) is ${\bm S}_0$, then its value at
the drain contact ($x=L$) can be found as $\left.{\bm
S}\right|_{t=L/v_0}$. Of course, the spin state at the drain
depends on the phase of ac field. For simplicity, we adopt the
value averaged over this parameter.

In the frame of reference rotating around ${\bm \omega}_0$ with
the frequency $\omega$, it is well known that electron spin
dynamics can be described as precessing around the vector
$(\omega_0-\omega,\omega_1/2; 0)$ if no relaxation is present (the
factor $1/2$ is due to the fact that we consider a linearly
polarized ac field). Effective resonance control of the spin
polarization occurs if the spin at the source is parallel to
${\bm\omega}_0$. If there is no resonance of $\omega_0$ and
$\omega$, then the spin remains practically unchanged during the
transfer to the drain. In contrast, under the resonance condition
($\omega \approx \omega_0$), spin precesses around ${\bm
\omega_1}$ with the frequency $\omega_1/2$ in the rotating frame.
In particular, if $\omega_1 L/v_0 =2\pi$, the spin flips during
the transfer through the QW, which can be registered by the drain
current. This clearly provides a possibility to probe the
frequency of the microwave signal by a simple electrical tuning of
the system into resonance.

Let us discuss the necessary conditions for supporting effective
resonance in the structure. Presumably, spin diffusion makes the
resonance weaker. Therefore, the drift term in Eq.~(\ref{eq:2})
must exceed the diffusion term. The characteristic length of spin
change is $v_0/\omega_0$. Using the Einstein relation for the
diffusion coefficient, we find that the diffusion can be
neglected if $(\Omega(\bar{k}))^2 \tau_p /\omega_0 \ll 1$, where
$\tau_p$ is the electron momentum relaxation time which determines
the mobility. This condition means that it is desirable to apply a
sufficiently strong dc electric field in a low-mobility sample. The
second condition requires that many precession cycles must occur
during the spin transfer through the QW. Therefore the QW length
must be $L \gg L_0 \equiv \hbar /(m \alpha)$, where
$L_0$ depends only on the material parameters. Finally,
the spin decay during the transfer must be negligible: $(\gamma
/\omega_0) (L/L_0) \ll 1$.

We can estimate the amplitude of the ac field $E^*$ required for
spin flip. Using the estimate for spin relaxation rate, we obtain
$e E^*\gg \hbar \alpha \bar{k}^2 (\tau/\tau_p)$. Clearly, the
sensitivity of the device is better for materials with a weaker SO
interaction. This is not surprising since the SO coupling controls
both spin precession and relaxation. Note that for a wide range of
structure parameters, $\tau$ is determined by the
electron-electron collisions \cite{ee1,ee2}  and $\tau/\tau_p \ll
1$.

Figure~1 shows the numerical results for the ratio $S_d/S_s$,
where $ S_s$ and $S_{d}$ are the $x$ component of the spin at the
source and drain contacts, respectively.  In the following, we
will refer to this ratio as {\em spin response.\/} The calculation
is performed for the microwave radiation frequency of 50~GHz in a
GaAs/AlGaAs QW with $\alpha = 5\times10^5~cm/s$, $\gamma = 6\times
10^9~s^{-1}$, the mobility $\mu = 2500~cm^{2}/Vs$, $L=10~\mu m$,
and several values of ac field $E_1$. The parameters used in this
study are close to those of the GaAs QW measured at room
temperature.\cite{exp}

As can be seen from the figure, the resonance in the spin response
appears for a relatively large $E_1$ (note that for a
electromagnetic plane wave, $E_1 = 1~kV/cm$ corresponds roughly to
the energy flux of $1.5~kW/cm^2$ in vacuum). For a moderately high
$E_1$, the resonance dip of the spin response increases with
$E_1$. This corresponds to the gradual increase of the spin
rotation angle up to a complete flip. The further increase of
$E_1$ leads to a nonresonant spin response character due to
multiple spin rotation cycles during the electron transfer between
the source and the drain. However, even for a strong field the
resonant character of spin response can be restored by changing
the device orientation with respect to the ac-field polarization.
This is because the spin rotation is mainly due to the ac-field
component perpendicular to the dc field. Such a behavior is
illustrated in Fig.~2, where the spin response at $E_1=1.5~kV/cm$
is plotted  for different angles $\phi$ between the ac and dc
fields; clearly, the $\phi= \pi/8$ case recovers a pronounced
resonance pattern.  In the inset, we show the dependence of spin
response on $E_1$ under the resonance condition
($\omega=\omega_0$). As expected, a periodic dependence is
observed, which corresponds to the periodic change in the phase of
the rotating spin at the drain. If the dc and ac fields are
parallel, the spin response becomes a smooth function of dc field.

The observed spin response properties suggest that the proposed
device structure can be used for detecting the frequency,
amplitude, and polarization of a microwave radiation. Under a
measurement set-up, it is possible first to detect the
polarization by finding device orientation characterized by the
absence of resonant features in the spin response (namely, the
orientation that the device dc field aligns with the microwave ac
field). Then, repositioning the device orientation for clear
resonant features, one can determine the frequency by the location
of resonance and the amplitude by its depth. Note that it is
possible to determine if the radiation is circularly polarized by
the change of the spin response when the dc bias is reversed.

Let us stress once more that scanning over the frequency range
occurs by simple variation of the dc field. The alternative
methods of microwave frequency measurements require either the
mechanical tuning of microwave cavity eigenfrequency or the use of
a superheterodyne technique, followed by intensity measurements
made by the devices insensitive to the frequency, such as
thermocouples, bolometers, or Schottky diodes.\cite{muwave} Since
these approaches allow detection of much weaker radiation than the
considered above, the proposed resonance method is most suitable
for high-power applications where measurements require the use of
massive calorimetric devices or attenuation
techniques.\cite{muwave} It is also important to note that the
proposed device practically does not absorb power; in the
structure considered with $E_1 =1~kV/cm$, the  absorbed ac power
density is about $8~W/cm^2$ for an electron density of $4\times
10^{10}~cm^{-2}$, which is more than two orders of magnitude less
than the microwave radiation energy flux. This means that the
device effectively does not disturb propagation of radiation.

Furthermore, the sensitivity of the proposed scheme may improve by
proper choice of the material system. Recently, weak SO
interaction was claimed in bulk GaN and GaN-based QWs
\cite{nitrides} with $\beta$ as low as $10^5~cm/s$. According to
the estimate, this provides a possibility to reduce the radiation
power requirement by more than an order of magnitude compared to
the GaAs-based counterpart. Alternatively, the measurement can be
achieved by the so-called electrically detected magnetic resonance
(EDMR) (see, for example, Refs.~\onlinecite{edmr1} and
\onlinecite{edmr2}). In this case, the frequency is determined by
the resonant change of differential conductivity due to electron
spin polarization and does not require spin-polarized electrodes.
Since the physical basis of EDMR remains unclear, it is difficult
to estimate the degree of conductivity variation for the
considered case of electrically-controlled resonance.

In conclusion, we predict an effective resonance control of
nonequilibrium electron spin by a microwave electric field. Based
on this phenomenon, it is possible to achieve efficient
electrically tuned measurement of frequency, polarization, and
amplitude of the microwave radiation.

The authors would like to thank V.I.~Sheka for fruitful
discussions. The work was supported by the Defence Advanced
Research Projects Agency and the CRDF (grant UE2-2439-KV-02).

\newpage
\begin{center}
Figure Captions
\end{center}

\vspace{12pt} \noindent FIG.~1. Spin response as a function of
effective precession frequency for a GaAs/AlGaAs QW for several
values of microwave field. The calculations are done for the
microwave frequency of 50~GHz. The insert shows the schematics of
the proposed device. The black regions represent spin-polarized
contacts connected by the QW.

\vspace{12pt} \noindent FIG.~2. Spin response for the amplitude of
ac field $E_1=1.5~kV/cm$ and different angles $\phi$ between the
ac and dc fields. The insert shows the dependence of spin response
on $E_1$ in resonance for two values of $\phi$.

\clearpage
\begin{figure}
\vspace{5.5cm}
\includegraphics[scale=0.9]{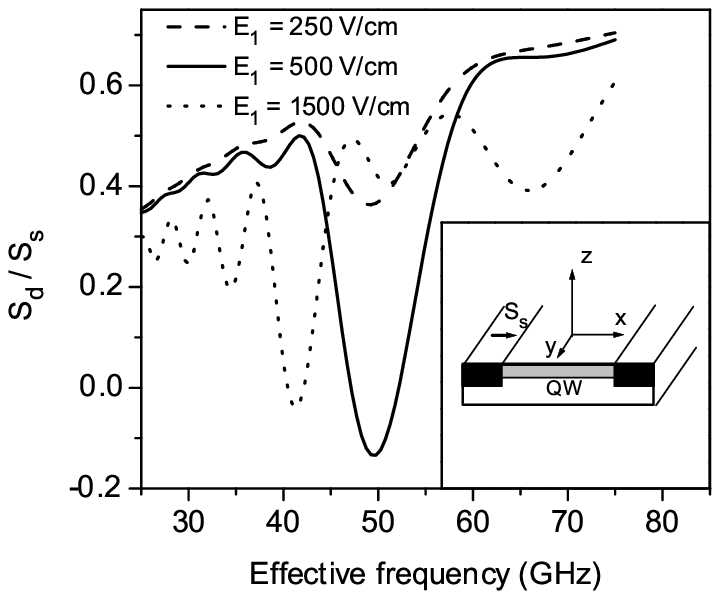}
\vspace{2in} \large Fig. 1.
\end{figure}

\newpage
\begin{figure}
\vspace{5.5cm}
\includegraphics[scale=0.9]{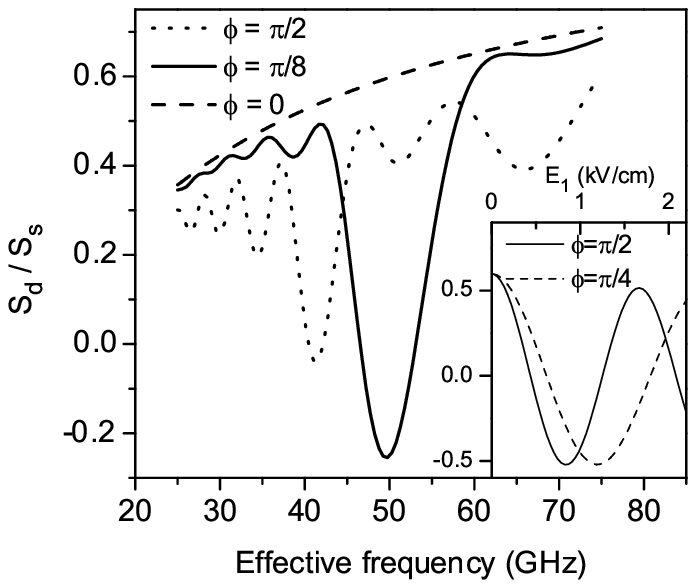}
\vspace{2in} \large Fig. 2.
\end{figure}


\newpage
\begin{thebibliography}{99}
%
\bibitem{dasdatta}
S.~Datta and B.~Das, Appl. Phys. Lett. {\bf 56}, 665 (1990).

\bibitem{byra}
Yu.~A.~Bychkov and E.~I.~Rashba, Pis'ma Zh. Eksp. Teor. Fiz. {\bf
39}, 66 (1984) [Sov. Phys. JETP Lett. {\bf 39}, 78 (1984)].

\bibitem{rashe}
E.~I.~Rashba and V.~I.~Sheka, in {\em Landau Level
Spectroscopy,\/} edited by G.~Landwehr and E.~I.~Rashba (Elsevier,
New York, 1991).

\bibitem{muwave}
G.~H.~Bryant, {\em Principles of Microwave Measurments\/} (Peter
Peregrinus, London, 1993).

\bibitem{previous}
A.~G.~Aronov and Yu.~B.~Lyanda-Geller, Pis'ma Zh. Eksp. Teor. Fiz.
{\bf 50} 398 (1989) [JETP Lett. {\bf 50}, 431 (1989)];
V.~M.~Edelstein, Solid State Commun. {\bf 73}, 233 (1990);
L.~I.~Magarill, A.~V.~Chaplik, and M.~V.~Entin, Fiz. Techn.
Polupr. {\bf 35}, 1128 (2001) [Semicond. {\bf 35}, 1081 (2001)].

\bibitem{wig}
S.~Saikin, cond-matt/0311221.

\bibitem{DP}
M.~I.~D'yakonov and V.~I.~Perel, Fiz. Tverd. Tela {\bf 13}, 3581
(1971) [Sov. Phys. Solid State {\bf 13}, 3023 (1972);
M.~I.~D'yakonov and V.~Yu.~Kachorovskii, Fiz. Techn. Polupr. {\bf
20}, 178 (1986) [Sov. Phys. Semicond. {\bf 20}, 110 (1986)].

\bibitem{abragam}
A.~Abragam, {\em The Principles of Nuclear Magnetism\/}
(Clarendon, Oxford, 1961).

\bibitem{ee1}
M.~M.~Glazov and E.~L.~Ivchenko, Pis'ma Zh. Eksp. Teor. Fiz. {\bf
75}, 476 (2002) [JETP Lett. {\bf 75}, 403 (2002)].

\bibitem{ee2}
R.~T.~Harley, M.~A.~Brand, A.~Malinowski, O.~Z.~Karimov,
P.~A.~Marsden, A.~J.~Shields, D.~Sanvitto, D.~A.~Ritchie, and
M.~Y.~Simmons, Physica E {\bf 17}, 324 (2003).

\bibitem{exp}
Y.~Ohno, R.~Terauchi, T.~Adachi, F.~Matsukura, and H.~Ohno,
Physica E {\bf 6}, 817 (2000).

\bibitem{nitrides}
S.~Krishnamurthy, M.~van~Schilfgaarde, and N.~Newman,
Appl. Phys. Lett. {\bf 83}, 1761 (2003); V.I.~Litvinov,
Phys. Rev. B {\bf 68}, 155314 (2003).

\bibitem{edmr1}
M.~Gu\`{e}ron and I.~Solomon, Phys. Rev. Lett. {\bf 15}, 667,
(1965); A.~Honig, Phys. Rev. Lett. {\bf 17}, 186 (1966); {\bf 17},
188 (1966).

\bibitem{edmr2}
C.F.O.~Graef, S.~Brandt, M.~Stutzmann, M.~Holzmann, G.~Abstreiter,
and F.~Sch\"{a}ffler, Phys. Rev. B {\bf 59}, 13242 (1999).

\end{thebibliography}
\end{document}